\documentclass[twocolumn,showpacs,preprintnumbers,superscriptaddress,amsmath,amssymb]{revtex4}
\usepackage{amsfonts}
\usepackage{amsmath}
\usepackage{amssymb}
\usepackage{graphicx}%
\usepackage{footmisc}
\begin{document}
\title{Spontaneous creation of the universe from nothing}
\author{Dongshan He}
\affiliation{State Key Laboratory of Magnetic Resonances and Atomic and Molecular Physics,
Wuhan Institute of Physics and Mathematics, Chinese Academy of Sciences, Wuhan
430071, China}
\affiliation{Graduate University of the Chinese Academy of Sciences, Beijing 100049, China}
\author{Dongfeng Gao}
\affiliation{State Key Laboratory of Magnetic Resonances and Atomic and Molecular Physics,
Wuhan Institute of Physics and Mathematics, Chinese Academy of Sciences, Wuhan
430071, China}
\author{Qing-yu Cai}
\thanks{Corresponding author. Electronic address: qycai@wipm.ac.cn}
\affiliation{State Key Laboratory of Magnetic Resonances and Atomic and Molecular Physics,
Wuhan Institute of Physics and Mathematics, Chinese Academy of Sciences, Wuhan
430071, China}

\begin{abstract}
An interesting idea is that the universe could be spontaneously created from
nothing, but no rigorous proof has been given. In this paper, we present such
a proof based on the analytic solutions of the Wheeler-DeWitt equation (WDWE).
Explicit solutions of the WDWE for the special operator ordering factor
$p=-2$ (or 4) show that, once a small true vacuum bubble is created by quantum
fluctuations of the metastable false vacuum, it can expand exponentially no
matter whether the bubble is closed, flat or open.
The exponential expansion
will end when the bubble becomes large and thus the early universe appears.
With the de Broglie-Bohm quantum trajectory theory, we show explicitly that it
is the quantum potential that plays the role of the cosmological constant and
provides the power for the exponential expansion of the true vacuum bubble. So
it is clear that the birth of the early universe completely depends on the
quantum nature of the theory.

\end{abstract}

\maketitle

\section{Introduction}

With the lambda-cold dark matter ($\Lambda$CDM) model
and all available observations (cosmic microwave background, abundance of
light elements), it has been widely accepted that the universe was created in
a big bang. However, there are still some puzzles, such as problems of the
flatness, the horizon, the monopole, and the singularity \cite{bb06}. Quantum
mechanics has been applied to cosmology to study the formation of the universe
and its early evolution. In particular, inflation theories, which suggest that
the universe experienced an exponential expansion period, were proposed to
solve puzzles of the early universe~\cite{as79,as80,ag81}. In quantum
cosmology theory, the universe is described by a wave function rather than the
classical spacetime. The wave function of the universe should satisfy the
Wheeler-DeWitt equation (WDWE)~\cite{bd67}. With the development of quantum
cosmology theory, it has been suggested that the universe can be created
spontaneously from nothing, where ``nothing" means there is neither matter nor
space or time \cite{av94}, and the problem of singularity can be avoided naturally.

Although the picture of the universe created spontaneously from nothing has
emerged for a long time, a rigorous mathematical foundation for such a picture
is still missing. According to Heisenberg's uncertainty principle, a small empty
space, also called a small true vacuum bubble, can be created
probabilistically by quantum fluctuations of the metastable false vacuum.
But if the small bubble cannot expand rapidly, it will disappear soon due to
quantum fluctuations. In this case, the early universe would disappear before
it grows up. On the other side, if the small bubble expands rapidly to a large
enough size, the universe can then be created irreversibly.

In this paper, we obtain analytic solutions of the WDWE of the true vacuum
bubble. With the de Broglie-Bohm quantum trajectory theory, we prove that once
a small true vacuum bubble is created, it has the chance to expand exponentially when it is
very small, \textit{i.e.}, $a\ll1$. The exponential expansion will end when
the true vacuum bubble becomes very large, \textit{i.e.}, $a\gg1$. It is the
quantum potential of the small true vacuum bubble that plays the role of
the cosmological constant and provides the power for its exponential
expansion. This explicitly shows that the universe can be created
spontaneously by virtue of a quantum mechanism.

\section{WDWE for the simplest minisuperspace model}

Heisenberg's
uncertainty principle indicates that a small true vacuum bubble can be created
probabilistically in a metastable false vacuum. The small bubble has 1
degree of freedom, the bubble radius. We can assume that the bubble is nearly
spherical, isotropic and homogeneous, since it is a small true vacuum bubble.
As we will show below, the small bubble will expand exponentially after its
birth and all asymmetries will be erased by the inflation.

Since the small true vacuum bubble is nearly spherical, it can be described by
a minisuperspace model \cite{npn00,npn12,apk97} with one single parameter of
the scalar factor $a$. The action of the minisuperspace can be written as
\begin{equation}
S=\frac{1}{16\pi G}\int R\sqrt{-g}d^{4}x. \label{action}%
\end{equation}
Since the bubble is homogeneous and isotropic, the metric in the
minisuperspace model is given by
\begin{equation}
ds^{2}=\sigma^{2}[N^{2}(t)dt^{2}-a^{2}(t)d\Omega_{3}^{2}]. \label{metric}%
\end{equation}
Here, $N(t)$ is an arbitrary lapse function, $d\Omega_{3}^{2}$ is the metric
on a unit three-sphere, and $\sigma^{2}=2G/3\pi$ is a normalizing factor
chosen for later convenience. Substituting Eq. (\ref{metric}) into Eq.
(\ref{action}), we obtain the Lagrangian
\begin{equation}
\mathcal{L=}\frac{N}{2}a(k-\frac{\dot{a}^{2}} {N^{2}}), \label{L}%
\end{equation}
where the dot denotes the derivative with respect to the time, $t$, and the
momentum
\begin{equation}
p_{a}=-a\dot{a}/N.\nonumber
\end{equation}
The Lagrangian (\ref{L}) can be expressed in the canonical form,
\begin{equation}
\mathcal{L=}p_{a}\dot{a}-N\mathcal{H},\nonumber
\end{equation}
where
\begin{equation}
\mathcal{H=-}\frac{1}{2}(\frac{p_{a}^{2}}{a}+ka).\nonumber
\end{equation}

In quantum cosmology theory, the evolution of the universe is completely
determined by its quantum state that should satisfy the WDWE. With
$\mathcal{H}\Psi=0$ and $p_{a}^{2}=-a^{-p} \frac{\partial}{\partial a}%
(a^{p}\frac{\partial}{\partial a})$, we get the WDWE \cite{swh84,av94}
\begin{equation}
(\frac{1}{a^{p}}\frac{\partial}{\partial a}a^{p}\frac{\partial}{\partial a}-k
a^{2})\psi(a)=0. \label{wdwm1}%
\end{equation}
Here, $k=1,0,-1$ are for spatially closed, flat, and open bubbles,
respectively. The factor $p$ represents the uncertainty in the choice of
operator ordering. For simplicity, we have set $\hbar=c=G=1$.

\section{Quantum trajectory from WDWE}

The complex function $\psi(a)$ can be
rewritten as
\begin{equation}
\psi(a)=R(a)\exp(iS(a)), \label{psi}%
\end{equation}
where $R$ and $S$ are real functions \cite{bd52,prh93}. Inserting $\psi(a)$
into Eq. (\ref{wdwm1}) and separating the equation into real and imaginary
parts, we get two equations \cite{bd52,prh93}:
\begin{align}
S^{\prime\prime}+2\frac{R^{\prime}S^{\prime}}{R}+\frac{p}{a}S^{\prime}  &  =0
,\label{wdwm2}\\
(S^{\prime})^{2}+V+Q  &  =0. \label{wdwm3}%
\end{align}
Here $V(a)=k a^{2}$ is the classical potential of the minisuperspace, the
prime denotes derivatives with respect to $a$, and $Q(a)$ is the quantum
potential,
\begin{equation}
Q(a)=-(\frac{R^{\prime\prime}}{R}+\frac{p}{a}\frac{R^{\prime}}{R}). \label{qp}%
\end{equation}

In the minisuperspace model, the current is~\cite{av88}
\begin{align}
j^{a}  &  =\frac{i}{2}a^{p}(\psi^{\ast}\partial_{a}\psi-\psi\partial_{a}
\psi^{\ast})\nonumber\\
&  =-a^{p}R^{2}S^{\prime}\nonumber
\end{align}
From Eq. (\ref{wdwm2}), we derive the following equations step by step:
\begin{align}
\frac{p}{a}RS^{\prime}+\frac{1}{R}(R^{2}S^{\prime})^{\prime}  &
=0,\nonumber\\
\frac{d(R^{2}S^{\prime})}{R^{2}S^{\prime}}+p\frac{da}{a}  &  =0,\nonumber\\
a^{p}R^{2}S^{\prime}  &  = \mathrm{const}.\nonumber
\end{align}
Then we have $\partial_{a}j^{a}=0$. This implies that Eq. (\ref{wdwm2}) is the
continuity equation.

It should be pointed out that Eq. (\ref{wdwm3}) is similar to the classical
Hamilton-Jacobi equation, supplemented by an extra term called quantum
potential $Q(a)$. $R$ and $S$ in Eq. (\ref{wdwm3}) can be obtained
conveniently from $\psi(a)$ by solving Eq. (\ref{wdwm1}) with relations,
\begin{align}
\psi(a)  &  =U+iW=R(a)\exp(iS(a)),\label{rs0}\\
R^{2}  &  =U^{2}+W^{2},\,\,\, S=\tan^{-1}(W/U). \label{rs}%
\end{align}
Generally speaking, the wave function of the bubble should be complex.
Specifically, if the wave function of the universe is pure real or pure imaginary
($W=0$ or $U=0$), we have $S^{\prime}=0$. That means the quantum potential $Q$
will counteract the ordinary potential $V$ at all times. Thus, the vacuum
bubble would evolve at a constant speed, and the small bubble cannot grow up
rapidly. In the following, we consider the general case for the vacuum bubble,
i.e., both $U$ and $W$ are nonzero functions.

By analogy with cases of nonrelativistic particle physics and quantum field
theory in flat space-time, quantum trajectories can be obtained from the
guidance relation~\cite{lpg93,npn00},
\begin{align}
\frac{\partial\mathcal{L}}{\partial\overset{\cdot}{a}}  &  =-a\dot{a}
=\frac{\partial S}{\partial a}\,,\label{gr}\\
\dot{a}  &  =-\frac{1}{a}\frac{\partial S}{\partial a} \,. \label{gr2}%
\end{align}
Equation (\ref{gr2}) is a first order differential equation, so the three-metric for
all values of the parameter $t$ can be obtained by integration.

\section{Inflation of the true vacuum bubble ($p\neq 1$)}

In the following, we solve the
WDWE of the bubble with $k=1,-1,0$, respectively. When the ordering factor
takes a special value $p=-2$ (or 4 for equivalence), exponential expansion
of the small true vacuum bubble induced by quantum potential can be obtained
no matter whether the bubble is closed, open, or flat.

\subsection{The closed bubble}

In this case, the analytic solution of
Eq. (\ref{wdwm1}) is
\begin{equation}
\psi(a)=a^{(1-p)/2}[ic_{1}I_{\nu}(\frac{a^{2}}{2})-c_{2}K_{\nu}(\frac{a^{2}%
}{2})], \label{psi1}
\end{equation}
where $I_{\nu}$'s are modified Bessel functions of the first kind,
$K_{\nu}$'s are the modified Bessel function of the second kind, the
coefficients $c_{1}$ and$\ c_{2}$ are arbitrary constants that should be
determined by the state of the bubble, and $\nu=|1-p|/4$.
As discussed
previously, the wave function of the bubble should be complex. For simplicity,
we set $c_{1}$ and$\ c_{2}$ as real numbers to find the inflation solution.

Using Eqs. (\ref{rs0}) and (\ref{rs}), we can get
\begin{equation}
S=\tan^{-1}[-\frac{c_{1}}{c_{2}}\frac{I_{\nu}(\frac{a^{2}}{2})}{K_{\nu}%
(\frac{a^{2}}{2})}], \nonumber\label{sp1}%
\end{equation}
and
\begin{equation}
R=a^{(1-p)/2}\sqrt{[c_{1}I_{\nu}(\frac{a^{2}}{2})]^{2}+[c_{2}K_{\nu}%
(\frac{a^{2}}{2})]^{2}}\,\,.\nonumber
\end{equation}
Here, we omit the sign \textquotedblleft$\pm$" in front of $R$, since it
does not affect the value of $Q(a)$ in Eq.(\ref{qp}). For small arguments
$0<x\ll\sqrt{\nu+1}$, Bessel functions take the following asymptotic forms:
\begin{equation}
I_{\nu}(x)\sim\frac{1}{\Gamma(\nu+1)}\left(\frac{x}{2}\right)^{\nu} \notag
\end{equation}
and
\begin{equation}
K_{\nu}(x)\sim\frac{\Gamma(\nu)}{2}\left(\frac{2}{x}\right)^{\nu}, \,\,\,\, \nu\neq0.\notag
\end{equation}
where $\Gamma(z)$ is the Gamma function.
It is easy to get
\begin{equation}
S(a\ll1)\approx-\frac{2c_{1}}{c_{2\Gamma(\nu)\Gamma(\nu+1)}}(\frac{a^{2}}%
{4})^{2\nu},\,\,\,\,\,\,\nu\neq0. \nonumber
\end{equation}
Using the guidance relation (\ref{gr2}), we can get
the general Bohmian trajectories for any small scale factor
%
%
$$ a(t)=\left\{
\begin{aligned}
&\left[\frac{(3-4\nu)\lambda(\nu)}{3}(t+t_{0})\right]^{\frac{1}{3-4\nu}
},\,\,\,\,\,\,\nu\neq 0, \,\, \frac{3}{4} \\
&e^{\lambda(3/4)(t+t_{0})},\,\,\,\,\,\,\,\,\,\,\,\,\,\,\,\,\,\,\,\,
\,\,\,\,\,\,\,\,\,\,\,\,\,\,\,\,\,\,\,\,\,\,\,\,\,\,\,\,\,\nu=\frac{3}
{4}, \\
\end{aligned}
\right.
$$
where $\lambda(\nu)=6c_{1}/(4^{2\nu}c_{2}\Gamma(\nu)\Gamma(\nu+1))$.
For the case of
$\nu=0$ ({\it i.e.}, $p=1$), we will discuss it later.

It is clear that only the ordering factor takes the value $p=-2$ (or $p=4$ for
equivalence), i.e., $\nu=3/4$, has the scale factor $a(t)$ an exponential behavior.
$\lambda(3/4)>0$ corresponds to an expansionary bubble,
and $\lambda(3/4)<0$ implies a contractive bubble that does not satisfy the
evolution of the early universe. Therefore, with the condition $\lambda(3/4)>0$, we
draw the conclusion that, for a closed true vacuum bubble, it can
expand exponentially, and then the early universe appears irreversibly.

The quantum mechanism of spontaneous creation of the early universe can be
seen from the quantum potential of the bubble. For the case of $p=-2$ (or 4),
the quantum potential of the small true vacuum bubble is
\begin{align}
Q(a\rightarrow0)=-a^{2}-\lambda(3/4)^2 a^{4}. \label{q1}
\end{align}
We find that the first term in quantum potential $Q(a\rightarrow0)$ exactly
cancels the classical potential $V(a)=a^{2}$. The effect of the second term
$-\lambda(3/4)^2 a^{4}$ is quite similar to that of the scalar field
potential in~\cite{hhh12} or the cosmological constant in~\cite{dhc05} for
inflation. For the small true vacuum bubble, we have $H\equiv\dot{a}/a$ and
$\Lambda=3H^{2}$. Then we can get the effective \textquotedblleft cosmological
constant" $\Lambda$ for the vacuum bubble as $\Lambda\approx3\lambda(3/4)^2$.
In this way, we can see that the quantum potential of the small true vacuum bubble
plays the role of the cosmological constant and provides the power for the
exponential expansion. It is the quantum mechanism (\textit{i.e.}, the quantum
potential) that dominates the exponential expansion of the vacuum bubble.

\subsection{The open bubble}

For this case, the analytic
solution of Eq. (\ref{wdwm1}) is found to be
\begin{equation}
\psi(a)=a^{(1-p)/2}\left[ic_{1}J_{\nu}\left(\frac{a^{2}}{2}\right)+c_{2}Y_{\nu}\left(\frac{a^{2}
}{2}\right)\right],\label{psi-1}
\end{equation}
where $J_{\nu}$'s are Bessel functions of the first kind, and $Y_{\nu}$'s
are Bessel function of the second kind and $\nu=|1-p|/4$. Likewise, we get
\begin{equation}
S=\tan^{-1}\left[\frac{c_{1}}{c_{2}}\frac{J_{\nu}(\frac{a^{2}}{2})}{Y_{\nu}
(\frac{a^{2}}{2})}\right],\nonumber
\end{equation}
and
\begin{equation}
R=a^{(1-p)/2}\sqrt{\left[c_{1}J_{\nu}\left(\frac{a^{2}}{2}\right)\right]^{2}+\left[c_{2}Y_{\nu}%
\left(\frac{a^{2}}{2}\right)\right]^{2}}\,\,.\notag
\end{equation}
For small arguments $0<x\ll\sqrt{\nu+1}$, Bessel functions take the
following asymptotic forms, $J_{\nu}(x)\sim
(x/2)^{\nu}/\Gamma(\nu+1)$, and $Y_{\nu}(x)\sim-\Gamma(\nu)
2^{\nu-1}/x^{\nu}$ for $(\nu\neq0)$.
Then we find
\begin{equation}
S(a\ll1)\approx-\frac{\pi c_{1}}{c_{2\Gamma(\nu)\Gamma(\nu+1)}}\left(\frac{a^{2}}
{4}\right)^{2\nu},\,\,\,\,v\neq0. \notag
\end{equation}
and
%
%
$$a(t)=\left\{
\begin{aligned}
& \left[\frac{(3-4\nu)\bar\lambda(\nu)}{3}(t+t_{0})\right]^{\frac{1}{3-4\nu}
}, \,\,\,\,\,\nu\neq 0, \,\, \frac{3}{4} \\
&e^{\bar\lambda(3/4)(t+t_{0})}, \,\,\,\,\,\,\,\,\,\,\,\,\,
\,\,\,\,\,\,\,\,\,\,\,\,\,\,\,\,\,\,\,\,\,\,\,\,\,\,\,\,\,\,\,\,\,\,\,
\nu=\frac{3}{4},
\end{aligned}
\right.
$$
where $\bar\lambda(\nu)=3\pi c_{1}/(4^{2\nu}c_{2}\Gamma(\nu)\Gamma
(\nu+1))$.

\bigskip It is interesting that the scale factor for the open bubble ($k=-1$)
is quite similar to that of the closed one ($k=1$). For the special case of
$p=-2$ (or 4), the scale factor $a(t)$
has an exponential behavior like before. In this case, the quantum potential
for the open bubble can be obtained as
\begin{equation}
Q(a\rightarrow0)=a^{2}-\bar\lambda(3/4)^2 a^{4}. \label{q-1}
\end{equation}
Comparing with the case of the closed bubble, we find that the terms $a^{2}$
in quantum potential $Q(a\rightarrow0)$ and classical potential $V(a)$ change
sign simultaneously, so they can still cancel each other exactly. Thus, it is
the term $-\bar\lambda(3/4)^2 a^{4}$ in quantum potential $Q(a\rightarrow0)$ that
causes the exponential expansion of the vacuum bubble. Likewise, we can get
the effective cosmological constant for the small true
vacuum bubble, $\Lambda\approx3\bar\lambda(3/4)^2$.

\subsection{The flat bubble}

The analytic
solution of Eq. (\ref{wdwm1}) is
\begin{equation}
\psi(a)=ic_{1}\frac{a^{1-p}}{1-p}-c_{2}, \label{psi0}%
\end{equation}
where $p\neq1$, and hence
\begin{align}
S  &  =\tan^{-1}[-\frac{c_{1}}{c_{2}}\frac{a^{1-p}}{1-p}], \,\,\,\, p\neq 1,\notag\\
R  &  =\sqrt{c_{2}^{2}+(c_{1}\frac{a^{1-p}}{1-p})^{2}}, \,\,\,\, p\neq 1. \notag
\end{align}
Using the guidance relation (\ref{gr2}), we can get the general form of the Bohmian
trajectories as
%
$$a(t)=\left\{
\begin{aligned}
&\left[\frac{c_1}{c_2}(3-|1-p|)(t+t_0)\right]^{\frac{1}{3-|1-p|}}, \, |1-p|\neq 0,3, \\
& e^{c_1(t+t_0)/c_2},\,\,\,\,\,\,\,\,\,\,\,\,\,\,\,\,\,\,\,\,\,\,\,\,\,\,\,\,
\,\,\,\,\,\,\,\,\,\,\,\,\,\,\,\,\,\,\,\,\,\,\,\,\,\,\,\,
|1-p|=3.
\end{aligned}
\right.
$$

Likewise, only conditions $p=-2$ (or 4) and $c_1/c_2>0$ are satisfied,
will the small true vacuum bubble expand exponentially.
For the case of exponential expansion, the
quantum potential for the vacuum bubble can be obtained as $Q(a\rightarrow
0)=-(c_{1}/c_{2})^{2}a^{4}$, while the classical potential is $V(a)=0$. This
definitely indicates that quantum potential $Q(a)$ is the origin of
exponential expansion for the small true vacuum bubble. Similarly, we can get
the effective cosmological constant for the small true
vacuum bubble as $\Lambda\approx3(c_{1}/c_{2})^{2}$.

\section{The Bohmian trajectories for $p=1$}

Solutions of Eq. (\ref{wdwm1})
for the $p=1$ case are still Eq. (\ref{psi1}) and Eq. (\ref{psi-1}) for
the closed and open bubbles, respectively.
For the flat bubble, the solution of Eq. (\ref{wdwm1})
for $p=1$ is
\begin{equation}
\psi(a)=ic_{1}-c_{2}\ln a. \notag
\end{equation}
It is clear that the quantum potential $Q(a)$ of the bubble approaches infinity when the bubble
is very small $a\rightarrow 0$, no matter whether the small bubble is closed,
open or flat. The requirement of a finite value of $Q(a\rightarrow0)$ will
result in $a(t)=\mathrm{constant}$ for $k=0,\pm1$.

\section{The behavior of large vacuum bubbles}

Let us look at behaviors of our solutions
for large vacuum bubbles \footnote{When $x\gg|{\nu}^{2}-1/4|$,
Bessel functions take the asymptotic forms,
$I_{\nu}(x)\sim e^{x}/\sqrt{2\pi x}$, $K_{\nu}(x)\sim e^{-x}
/\sqrt{2\pi x}$, $J_{\nu}(x\gg|{\nu}^{2}-1/4|)\sim\sqrt{2/\pi x}
\cos(x-\nu\pi/2-\pi/4)$, and $Y_{\nu}(x\gg|{\nu}^{2}-1/4|)\sim\sqrt{2/\pi x}\sin
(x-\nu\pi/2-\pi/4)$. We can use these asymptotic forms to
calculate $S(a)$ and $a(t)$ for large bubbles.}.
For the closed bubble, we get
$S(a\gg1)=-\tan^{-1}(c_1e^{a^2}/c_2)$, and hence $\dot{a}=2c_1e^{-a^2}/c_2\rightarrow 0$.
The quantum potential of the bubble is
$Q(a\gg1)\sim-a^{2}$. It is obvious that there is no classical limit for the
closed bubble.
For the open bubble, we have $S(a\gg1)\sim-\tan^{-1}[c_1\tan(a^{2}/2+\pi/4-\nu\pi/2)/c_2]$.
When $|c_1/c_2|=1$, we can get its classical limit $\dot{a}^2=1$ as $Q(a\gg 1)\rightarrow 0$.
For the case of the flat bubble, we get $\dot{a}=c_1a^{-|1-p|-2}/c_2$.
When the bubble becomes large enough, it can reach the classical limit,
$\dot{a}^2\rightarrow0$ with $Q(a\gg 1)\rightarrow 0$.
When the vacuum bubble becomes very large, it will stop expanding for $k=0,1$,
or it will expand with a constant velocity for $k=-1$.
In one word, it turns out that the vacuum bubble will stop accelerating when it becomes
very large, no matter whether it is closed, flat, or open.

\section{The operator ordering factor}

Generally speaking, the factor $p$ in Eq. (\ref{wdwm1}) represents the uncertainty of
the operator ordering. Different $p$ gives different rule of quantization for
the classical system. From Eqs. (\ref{wdwm2}) and (\ref{qp}), we get a
general form of the quantum potential,
\begin{equation}
Q(a)=-[\frac{-p^{2}+2p}{4a^{2}}+\frac{3(S^{\prime\prime})^{2}}{4(S^{\prime
})^{2}}-\frac{S^{\prime\prime\prime}}{2S^{\prime}}]. \label{qps}%
\end{equation}
It is clear that the effect of the ordering factor $p$ is important only to small
bubbles, and different $p$ will result in different quantum potential. In other words,
for small bubbles (i.e., $a\ll 1$), the first term is significant to $Q(a)$,
while for large bubbles (i.e., $a\gg1$), it is negligible. So, the factor $p$
represents quantum effects of the system
described by the WDWE in Eq. (\ref{wdwm1}).

It is interesting that only when the ordering factor $p$ takes value $-2$ (or
$4$) can one get the exponential expansion for the small true
vacuum bubble, no matter whether the bubble is closed, flat, or open.
It is generally believed that the operator ordering factor $p$ can be restricted
by the quantum to classical transition of the system \cite{hh83}.
Maybe a more elegant treatment of the quantum to classical transition
is needed to restrict the interesting values of $p$, since
the classical limit is independent of $p$ in the present treatment.
A hint from loop quantum gravity (LQG) theory is that when one wants to
remove the ambiguities from LQG, the ordering factor should take the value
$p=-2$ \cite{ns08}.

\section{Discussion and conclusion}

In summary, we have presented a
mathematical proof that the universe can be created spontaneously from
nothing. When a small true vacuum bubble is created by quantum fluctuations of
the metastable false vacuum, it can expand exponentially if the
ordering factor takes the value $p=-2$ (or 4). In this way, the
early universe appears irreversibly. We have shown that it is the quantum
potential that provides the power for the exponential expansion of the bubble.
Thus, we can conclude that the birth of the early universe is completely
determined by quantum mechanism.

One may ask the question when and how space, time and matter appear in the
early universe from nothing. With the exponential expansion of the bubble, it
is doubtless that space and time will emerge. Due to Heisenberg's
uncertainty principle, there should be virtual particle pairs created by
quantum fluctuations. Generally speaking, a virtual particle pair will
annihilate soon after its birth. But, two virtual particles from a pair can be
separated immediately before annihilation due to the exponential expansion of
the bubble. Therefore, there would be a large amount of real particles created
as vacuum bubble expands exponentially. A rigorous mathematical calculation
for the rate of particle creation with the exponential expansion of the bubble
will be studied in our future work.

\section{Acknowledgment}

We thank the referees for their helpful comments and suggestions that significantly
polish this work.
Financial support from NSFC under Grant No. 11074283, and NBRPC under Grant
Nos. 2013CB922003 is appreciated.

\end{document}